\documentclass[conference]{IEEEtran}
\IEEEoverridecommandlockouts
\usepackage{cite}
\usepackage{amsmath,amssymb,amsfonts}
\usepackage{algorithmic}
\usepackage{graphicx}
\usepackage{textcomp}
\usepackage{xcolor}
\usepackage{orcidlink}
\usepackage{multirow}
\usepackage{makecell}
\usepackage{array}
\usepackage[capitalize]{cleveref}

\crefname{equation}{}{}
\crefname{section}{Sec.}{Secs.}
\crefname{figure}{Fig.}{Figs.}
\Crefname{table}{Table}{Tabs.}

\def\BibTeX{{\rm B\kern-.05em{\sc i\kern-.025em b}\kern-.08em
    T\kern-.1667em\lower.7ex\hbox{E}\kern-.125emX}}
    
\makeatletter
\newcommand{\linebreakand}{%
  \end{@IEEEauthorhalign}
  \hfill\mbox{}\par
  \mbox{}\hfill\begin{@IEEEauthorhalign}
}
\makeatother

\begin{document}

\title{Lightweight Error-Correction Code Encoders in Superconducting Electronic Systems\\
\thanks{This work is supported in part by the National Science Foundation Expeditions Award under Grant CCF-2124453 and SHF Award under Grant CCF-2308863, and Department of Energy EXPRESS program under Grant DE-SC0024198.}
}

\author{
\IEEEauthorblockN{Yerzhan Mustafa}
\IEEEauthorblockA{\textit{Department of Electrical and Computer Engineering}\\
\textit{University of Rochester}\\
Rochester, NY, United States\\
yerzhan.mustafa@rochester.edu}
\and
\IEEEauthorblockN{Berker Peköz
\orcidlink{0000-0002-7572-3663}
}
\IEEEauthorblockA{\textit{Department of Electrical Engineering and Computer Science}\\
\textit{Embry-Riddle Aeronautical University}\\
Daytona Beach, FL, United States\\
pekozb@erau.edu}
\linebreakand 
\IEEEauthorblockN{Selçuk Köse}
\IEEEauthorblockA{\textit{Department of Electrical and Computer Engineering}\\
\textit{University of Rochester}\\
Rochester, NY, United States\\
selcuk.kose@rochester.edu}
}

\maketitle

\begin{abstract}
Data transmission from superconducting electronic circuits, such as single flux quantum (SFQ) logic, to room-temperature electronics is susceptible to bit errors, which may result from flux trapping, fabrication defects, and process parameter variations (PPV). 
Due to the cooling power budget at 4.2 K and constraints on the chip area, the size of the error-correction code encoders is limited. 
In this work, three lightweight error-correction code encoders are proposed that are based on Hamming(7,4), Hamming(8,4), and Reed-Muller(1,3) codes and implemented with SFQ logic. 
The performance of these encoders is analyzed in the presence of PPV. 
The trade-offs between the theoretical complexity and physical size of error-correction code encoders are identified.

\end{abstract}


\section{Introduction}
Superconducting digital electronics such as single flux quantum (SFQ) logic can operate at extremely high switching frequencies (tens to hundreds of GHz) and consume significantly low energy per switching activity, in the order of 10$^{-19}$~J \cite{likharev1991rsfq,krylov2024single}. 
SFQ logic technology is a promising candidate for beyond-CMOS technology, especially for large-scale applications such as data centers and cloud computing. 
Additionally, SFQ circuits can be used in large-scale in-fridge control and readout circuitry of superconducting quantum computers \cite{mukhanov2019scalable,jokar2022digiq}. 

In SFQ logic, the information is represented by the presence and absence of voltage pulses that correspond to the logical `1' and `0', respectively. 
These pulses are generated and transmitted by switching Josephson junctions (JJs), which are two terminal devices that consist of two superconductor materials separated by an insulator. 
The amplitude of the voltage pulse is around 1~mV with 2~ps duration. 
To interface SFQ circuits with room-temperature electronics (typically CMOS technology), SFQ pulses are amplified and converted to DC voltages - up to 1~V - by specialized superconducting output drivers and semiconductor amplifiers~\cite{gupta2019digital,ortlepp2013design,mustafa2024ternary,mustafa2025pam4_squid}. 

Data transmission from an SFQ chip to a higher temperature stage is subject to bit errors due to, \textit{e.g.}, flux trapping \cite{robertazzi1997flux,fourie2021experimental}, fabrication defects, and process parameter variations (PPV) \cite{tolpygo2014inductance}. 
SFQ circuits are therefore often designed to account for the circuit parameter variations up to $\pm20$ to $\pm30$\% of the nominal values \cite{mustafa2022optimization,mustafa2023suzuki}. 

An SFQ-based error-correction code encoder has been presented in \cite{peng2019solution}. It is based on a (38,32) linear block code and has a 32-bit input message and six parity bits. The (38,32) linear block code can detect 2-bit and correct 1-bit errors using a  circuit consisting of 84 XOR gates and 135 D flip-flops (DFFs) implemented with SFQ logic \cite{peng2019solution}. 
Due to the low integration density of superconducting circuits, the physical realization of SFQ-based processors is often limited to an 8-bit architecture \cite{ishida2018towards,ando2016design,qu2020design,nagaoka202257}. 
Additionally, the complexity of SFQ circuits is limited by the number of input/output/bias pins (\textit{e.g.}, 40~pins for a 5x5~mm$^2$ chip) and the heat load of cryogenic cables, which connect the thermal zones of the cryostat \cite{mustafa2024built,krinner2019engineering,mustafa2024dc,krause2024signal}. 
Due to these unique challenges of SFQ circuits, circuit-level mitigation strategies to address bit errors should minimize additional cable requirements and circuit area overhead.

\begin{figure*}[t]
	\centering
	\includegraphics[width=0.98\textwidth]{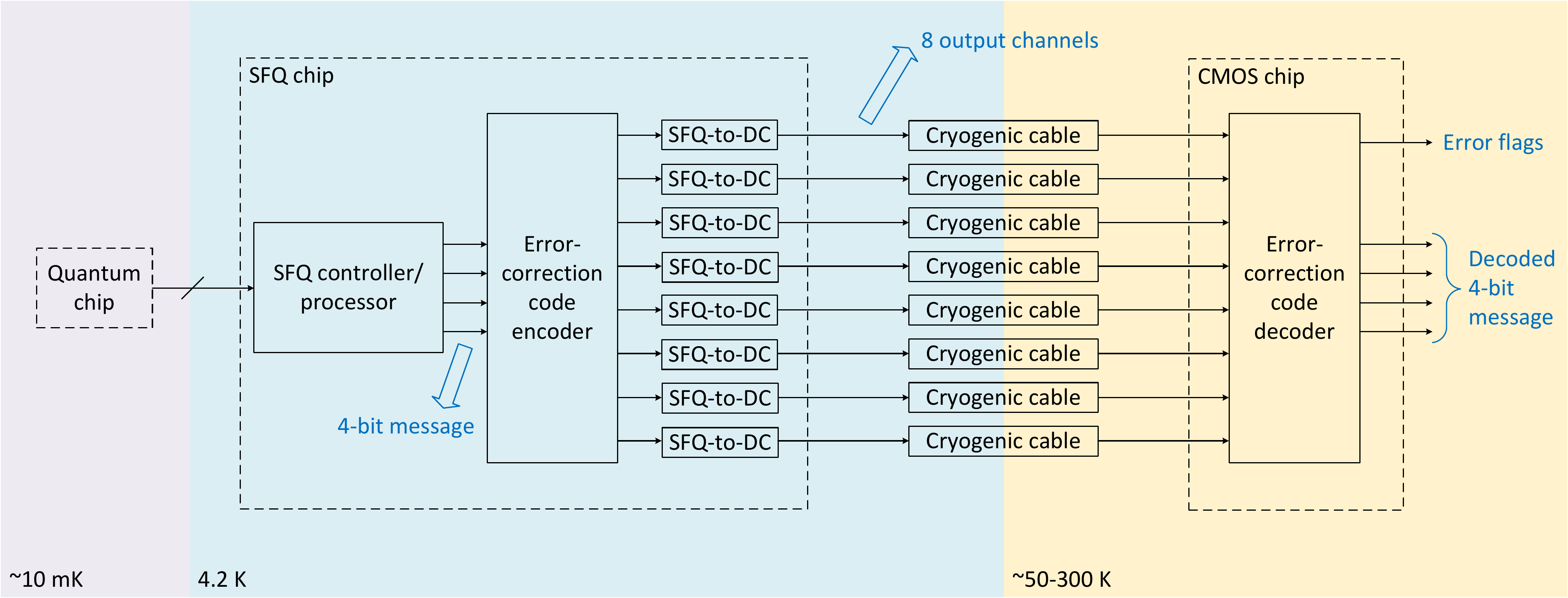}
	\caption{Block diagram of a cryogenic digital output data link incorporating an error-correction code encoder and decoder. CMOS amplifier circuits (not shown) may be included on the CMOS chip to boost the amplitude of the received signals.}
	\label{fig:cryogenic_encoder_decoder_block_diagram}
\end{figure*}

In this work, several lightweight error-correction code encoders are designed and compared. 
Due to the aforementioned practical limitations, our analysis will be limited to an 8-bit interface (output channels) and a 4-bit message that is transmitted from 4.2~K to a higher temperature stage (50-300~K), as shown in \cref{fig:cryogenic_encoder_decoder_block_diagram}.

The remainder of the paper is organized as follows. 
Lightweight error-correction codes such as Hamming and Reed-Muller are presented in Section \ref{section:ECC_overview}. 
The circuit-level implementation of these encoders is explained in Section \ref{section:ECC_circuit_design} and evaluated in Section \ref{section:ECC_evaluation}. 
Conclusions are drawn in Section~\ref{conclusion}.

\section{Lightweight Error-Correction Codes}\label{section:ECC_overview}
An SFQ processor that outputs 4-bit messages is illustrated in \cref{fig:cryogenic_encoder_decoder_block_diagram}. 
The information theory community has historically prioritized devising capacity achieving codes at asymptotic message lengths \cite{reed1960polynomial,gallager1962ldpc,berrou1993turbo,arikan2009polar} that require computationally intensive soft decision decoding \cite{gallager1962ldpc}, or resource-intensive methods such as using successive cancellation list decoding augmented with cyclic redundancy checks \cite{arikan2009polar} to approach theoretical performance limits.

In contrast, mission-critical embedded systems, particularly those operating under stringent latency, power, and hardware constraints (\textit{e.g.}, superconducting logic), demand lightweight error-correcting codes optimized for short blocklengths. 
In this regime, the Hamming and Reed-Muller codes offer compelling trade-offs between reliability and implementation complexity. 
Although Bose–Chaudhuri–Hocquenghem (BCH) codes \cite{hocquenghem1959codes,bose1960class} are algebraically equivalent to Hamming codes at short lengths, their higher encoding and decoding complexity makes them less suitable for resource-constrained environments.

\subsection{Hamming Codes}
Hamming codes, introduced by Richard Hamming in 1950 \cite{hamming1950error}, represent the first known class of nontrivial, scalable, and perfect single-error-correcting codes\cite{tietavainen_nonexistence_1973}.
These codes have low decoding complexity using the syndrome decoding concept introduced by Hamming, which points to the position in error when calculated, allowing correction by flipping the identified bit.

To enhance error detection capabilities, the original (7,4) Hamming code can be extended by appending an overall parity bit, yielding the quasi-perfect (8,4,4) extended Hamming code, which will be referred to as Hamming(8,4) in the rest of this work for the purpose of brevity. 
This extension increases the minimum distance $d_{min}$ from 3 to 4, enabling reliable detection of all 2- and 3-bit errors, while preserving single-error correction.

\subsection{Reed-Muller Codes}
Irving Reed\cite{reed1954class} and David Muller\cite{muller1954application} independently but simultaneously introduced Reed-Muller codes in 1954. Plotkin then introduced a construction method\cite{plotkin1960binary} that facilitates efficient encoding and decoding\cite{beery1986optimal}. The recursive nature not only enables simpler scalable hardware implementation, but also provides the ability to correct certain 2-bit error patterns~\cite{yasunaga2010correctable}.

\subsection{Comparison}
A comparative analysis of these lightweight codes is presented in \cref{tab:lwc}. 
For Hamming(7,4), the worst-case scenario arises when the decoder attempts correction and misclassifies 2- and 3-bit errors as a correctable 1-bit error and valid codeword, respectively, leading to undetected miscorrection, while it can correctly identify 28 out of the 35 possible 3-bit error patterns, an 80\% detection rate.

\begin{table}
\caption{Number of detected and corrected errors.}
    \centering
    \begin{tabular}{m{4em} m{2em} m{4em} m{4em} m{4em} m{4em}}
            & & \multicolumn{2}{|c|}{Worst case} & \multicolumn{2}{|c|}{Best case} \\
         Code & $d_{min}$ & Errors detected & Errors corrected & Errors detected & Errors corrected\\
         \hline\hline
         Hamming (7,4) & 3 & 1 & 1 & 3 & 1 \\
         \hline
        Hamming (8,4) & 4 & 3 & 1 & 3 & 1 \\
         \hline
         RM (1,3) & 4 & 3 & 1 & 3 & 2 \\
    \end{tabular}
    \label{tab:lwc}
\end{table}

\section{Circuit-Level Implementation of Encoders}\label{section:ECC_circuit_design}
Three different error-correction code encoders are designed using SFQ logic. 
SFQ circuits have two unique features that are not present in standard CMOS circuit design. 
First, all SFQ logic gates, such as AND, OR, XOR, and NOT gates, require a clock signal to generate an output signal \cite{likharev1991rsfq}.
Due to the requirement for a clock signal, data paths must be balanced to ensure proper timing alignment, which is typically achieved by adding D flip-flop (DFF) cells \cite{pasandi2018pbmap}. 
Second, the SFQ logic gates have a fan-out of one. 
An SFQ splitter circuit is therefore needed to drive two or more subsequent logic cells \cite{likharev1991rsfq}. 

As an example, let us consider the design of a Hamming(8,4) code encoder circuit. 
The generator matrix ($G$) of Hamming(8,4) code is given by
\begin{equation}
    G_{Hamming(8,4)} = 
\begin{bmatrix}
    1 & 1 & 1 & 0 & 0 & 0 & 0 & 1\\
    1 & 0 & 0 & 1 & 1 & 0 & 0 & 1\\
    0 & 1 & 0 & 1 & 0 & 1 & 0 & 1\\
    1 & 1 & 0 & 1 & 0 & 0 & 1 & 0
\end{bmatrix}.
\label{eq:generator_matrix_Hamming84}
\end{equation}
To generate the codeword, a message bit string should be multiplied by $G$ with$\mod2$ operator as
\begin{equation}
    codeword = (message\times G)\mod 2,
\label{eq:codeword_equation}
\end{equation}
\noindent
where $message=[m_1,m_2,m_3,m_4]$ is a 4-bit message and $codeword = [c_1,c_2,...,c_8]$ is an 8-bit codeword. 
The codeword can then be presented in boolean form as
\begin{equation}
\begin{array}{l}
     c_1 = m_1 \oplus m_2 \oplus m_4;\\
    c_2 = m_1 \oplus m_3 \oplus m_4;\\
    c_3 = m_1;\\
    c_4 = m_2 \oplus m_3 \oplus m_4;\\
    c_5 = m_2;\\
    c_6 = m_3;\\
    c_7 = m_4;\\
    c_8 = m_1 \oplus m_2 \oplus m_3,
\end{array}
\label{eq:codeword_boolean}
\end{equation}
\noindent
where $\oplus$ is an XOR operator. 

\begin{figure}[t]
	\centering
	\includegraphics[width=0.48\textwidth]{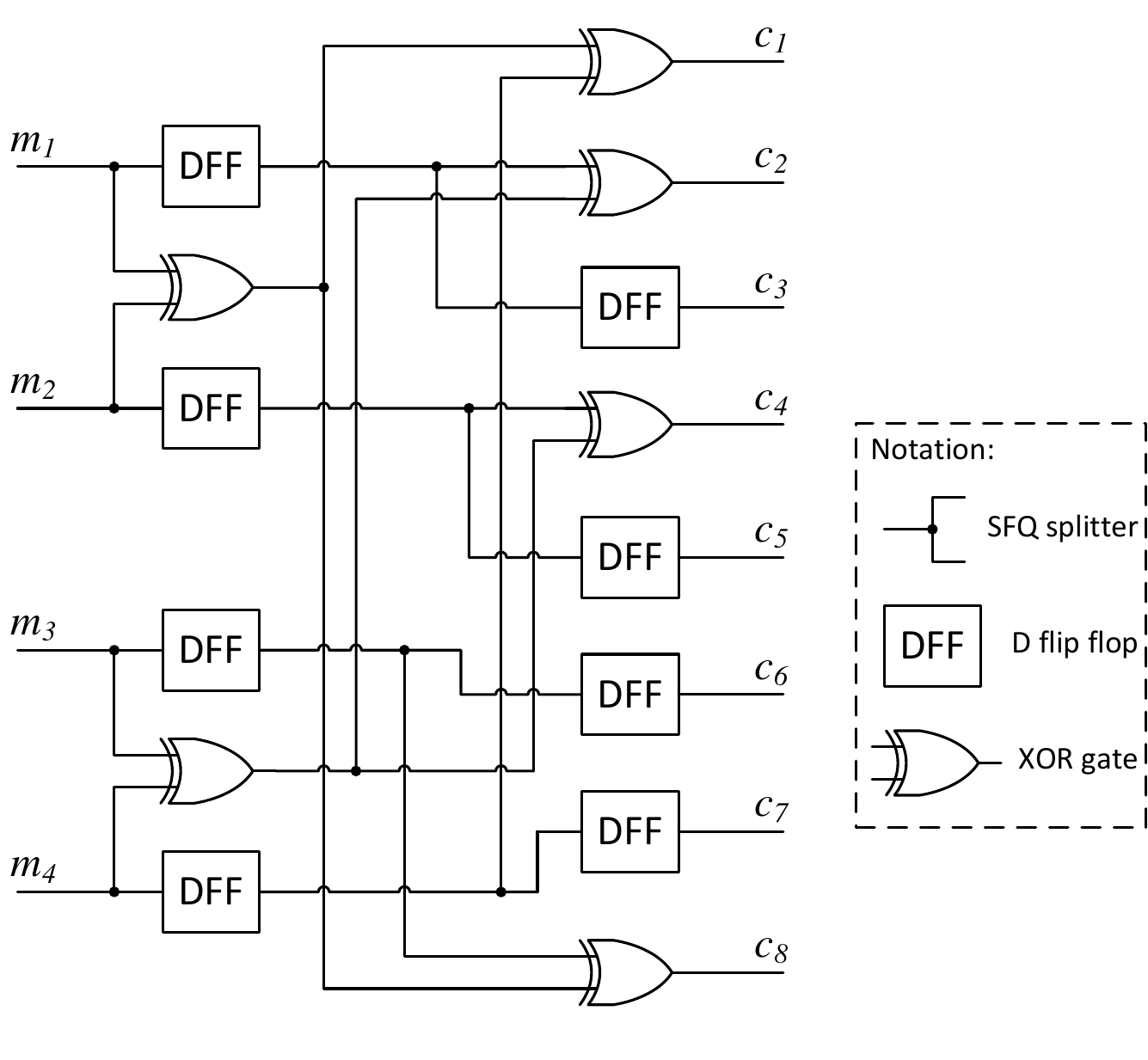}
	\caption{Schematic of a Hamming(8,4) code encoder implemented with SFQ logic. All XOR and DFF cells are clocked, though clock lines are not shown.}
	\label{fig:Hamming84_encoder_schematic}
\end{figure}

The schematic of a Hamming(8,4) code encoder circuit with SFQ logic cells is depicted in Fig.~\ref{fig:Hamming84_encoder_schematic}. 
The logic depth is equal to two, which is determined by $c_1,c_2,c_4,$ and $c_8$ in (\ref{eq:codeword_boolean}). 
Therefore, it takes two clock cycles to produce these codeword bits. 
To balance the arrival of remaining codeword bits (\textit{i.e.}, $c_3,c_5-c_7$), two DFFs are added for each path, as shown in Fig.~\ref{fig:Hamming84_encoder_schematic}.

\begin{figure}[t]
	\centering
	\includegraphics[width=0.49\textwidth]{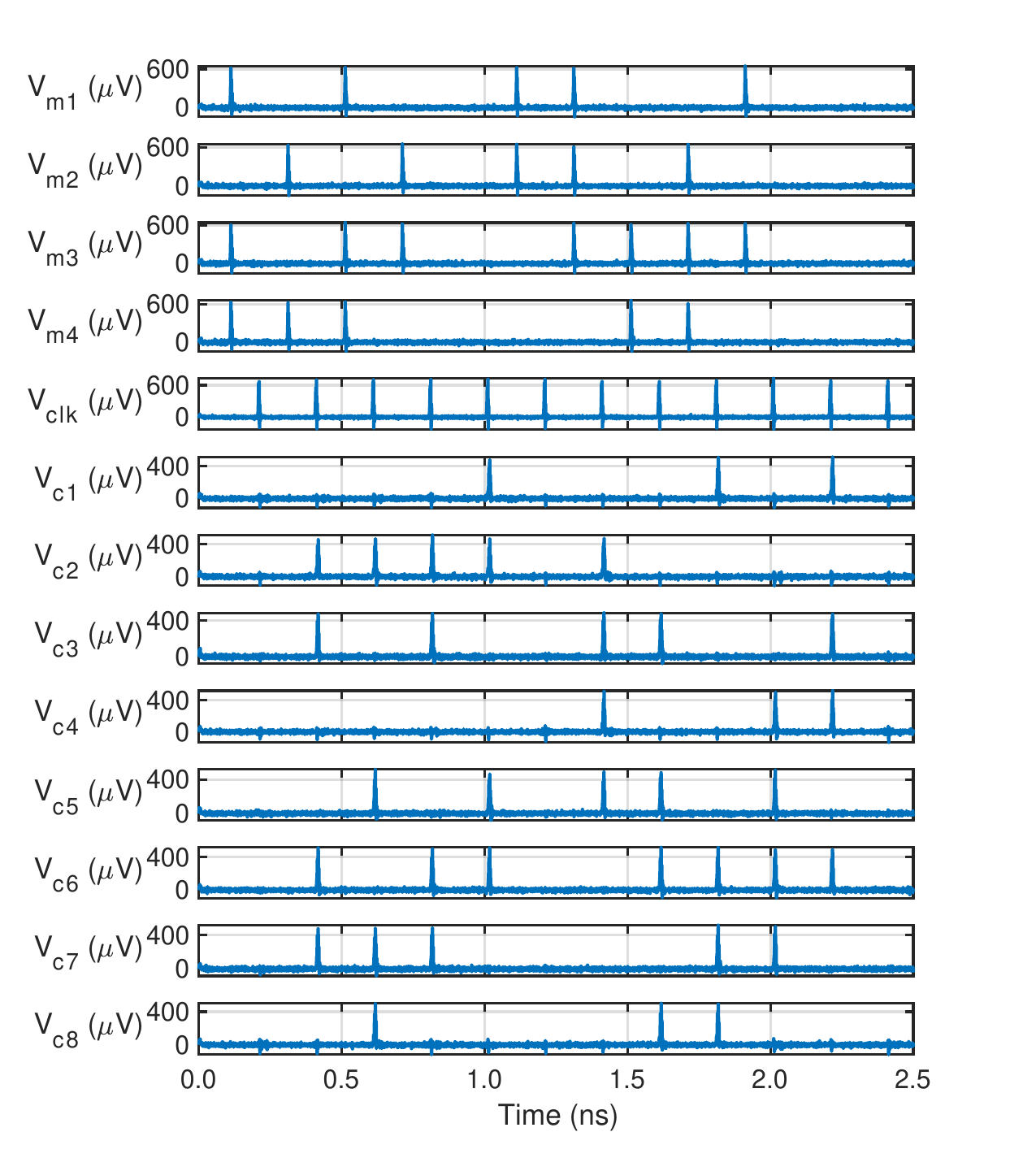}
	\caption{Simulation results of a Hamming(8,4) code encoder operating at 5~GHz. Thermal noise at 4.2 K is added.}
	\label{fig:Hamming84_waveforms}
\end{figure}

The simulation results of the Hamming(8,4) code encoder are shown in \cref{fig:Hamming84_waveforms}. 
The circuit is designed using  SuperTools/ColdFlux RSFQ cell library \cite{supertools_rsfq_cell_library} with MIT Lincoln Lab SFQ5ee 10~kA/cm$^2$ process. 
JoSIM, a superconductor SPICE tool \cite{delport2019josim}, is used as a simulator. 
As shown in \cref{fig:Hamming84_waveforms}, the codeword bits are produced after two clock cycles. For example, the message `1011' is applied at around 0.1 ns, and the corresponding codeword `01100110' is produced at 0.4 ns. Note that in SFQ logic, the presence and absence of a voltage pulse represent the logical `1' and `0', respectively.

\begin{figure}[t]
	\centering
	\includegraphics[width=0.48\textwidth]{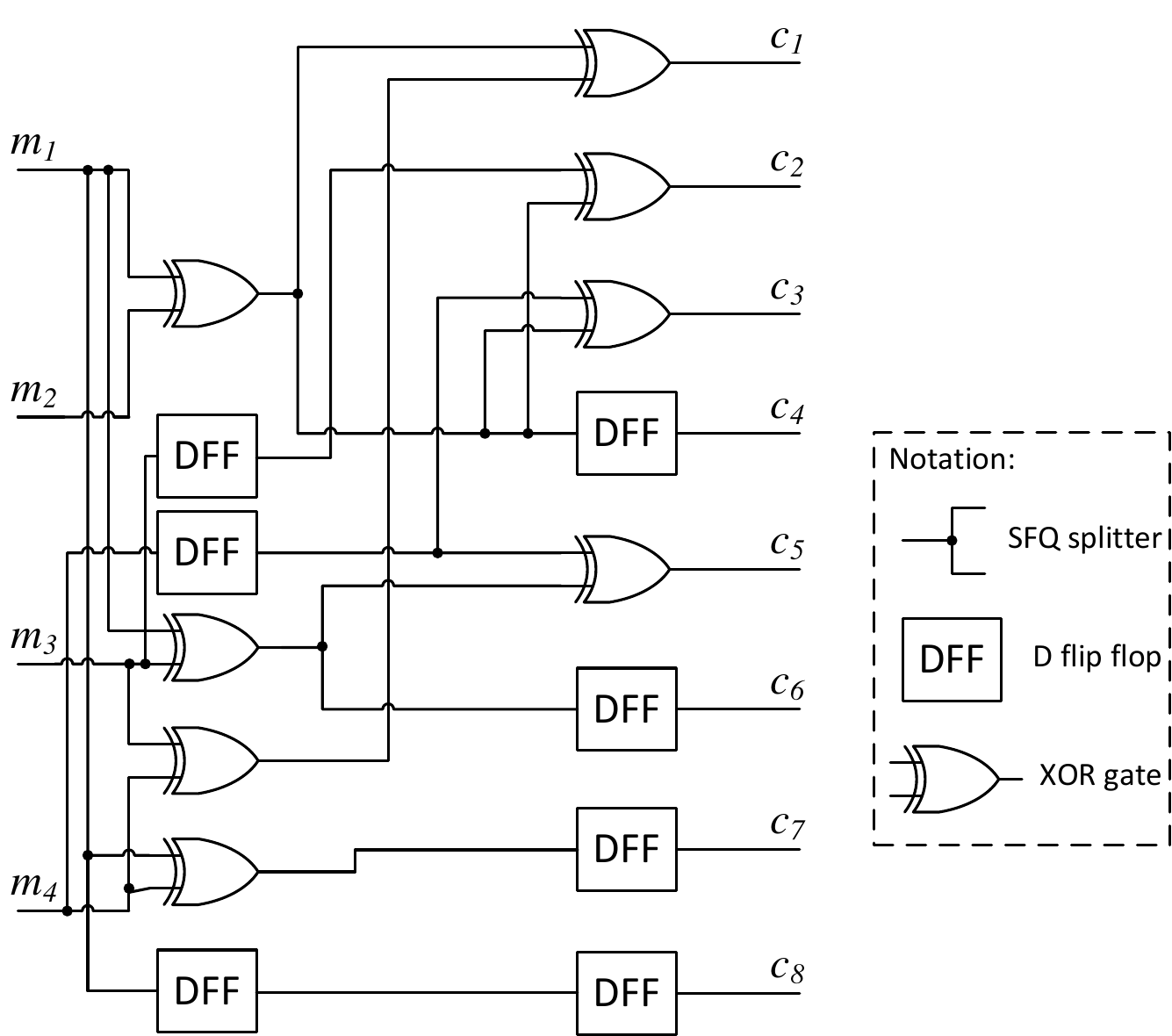}
	\caption{Schematic of an RM(1,3) code encoder implemented with SFQ logic. All XOR and DFF cells are clocked (not shown).}
	\label{fig:RM13_encoder_schematic}
\end{figure}

Following the same procedure, the Hamming(7,4) and RM(1,3) code encoder circuits are also designed. 
The schematic of the Hamming(7,4) code encoder circuit is similar to that of the Hamming(8,4) encoder without the output bit $c_8$. 
The schematic of RM(1,3) code encoder is shown in \cref{fig:RM13_encoder_schematic}.
The circuit level details of these encoders are listed in \cref{table:power_area_comparison}. 
It should be noted that in addition to, \textit{e.g.}, 10 SFQ splitters in Hamming(8,4) code encoder (\cref{fig:Hamming84_encoder_schematic}), 13 more splitters are needed to form a clock distribution network for XOR and DFF cells. \cref{table:power_area_comparison} lists the number of JJs, static power dissipation, and layout area of the encoders implemented with SuperTools/ColdFlux RSFQ standard cells \cite{supertools_rsfq_cell_library}. 

\begin{table*}[t]
\caption{Circuit-level comparison of error-correction code encoders.}
\begin{center}
\begin{tabular}{|c|c|c|c|c|}
        \hline
		\textbf{Encoder} & \makecell{\textbf{Standard} \textbf{cells}} & \makecell{\textbf{JJ} \textbf{count}} & \makecell{\textbf{Power} \textbf{dissipation ($\mu$W)}} & \makecell{\textbf{Layout} \textbf{area (mm$^2$)}}\\
		\hline
		Reed-Muller RM(1,3) & \makecell{8 XOR gates, \\7 DFFs, 26 splitters, \\8 SFQ-to-DC converters} & 305 & 101.5 & 0.193 \\
            \hline
            Hamming(7,4) & \makecell{5 XOR gates, \\8 DFFs, 20 splitters, \\7 SFQ-to-DC converters} & 247 & 81.7 & 0.158 \\
            \hline
            Hamming(8,4) & \makecell{6 XOR gates, \\8 DFFs, 23 splitters, \\8 SFQ-to-DC converters} & 278 & 92.3 & 0.177 \\
            \hline
\end{tabular}
\label{table:power_area_comparison}
\end{center}
    
\end{table*}


\section{Performance Evaluation of Encoders}\label{section:ECC_evaluation}
The performance of the aforementioned three error-correction code encoders is analyzed and compared in the presence of PPV. 
The PPV effect can be modeled in JoSIM using a `spread' function, where each circuit parameter (such as the critical current of JJs, inductance, and resistance) is assigned a specified deviation from the nominal parameter value. 
This deviation is typically the result of the imperfections in the fabrication process.

JoSIM SPICE simulator and MATLAB tools have been used to perform the performance analysis. 
A 4-bit random message is generated with a MATLAB script and is fed to the JoSIM netlist. 
Once the circuit-level implementation of error-correction code encoder is simulated in JoSIM, the output voltage waveforms are processed with MATLAB for signal decoding using standard error-correction code decoding techniques. 

\begin{figure}[t]
	\centering
	\includegraphics[width=0.5\textwidth]{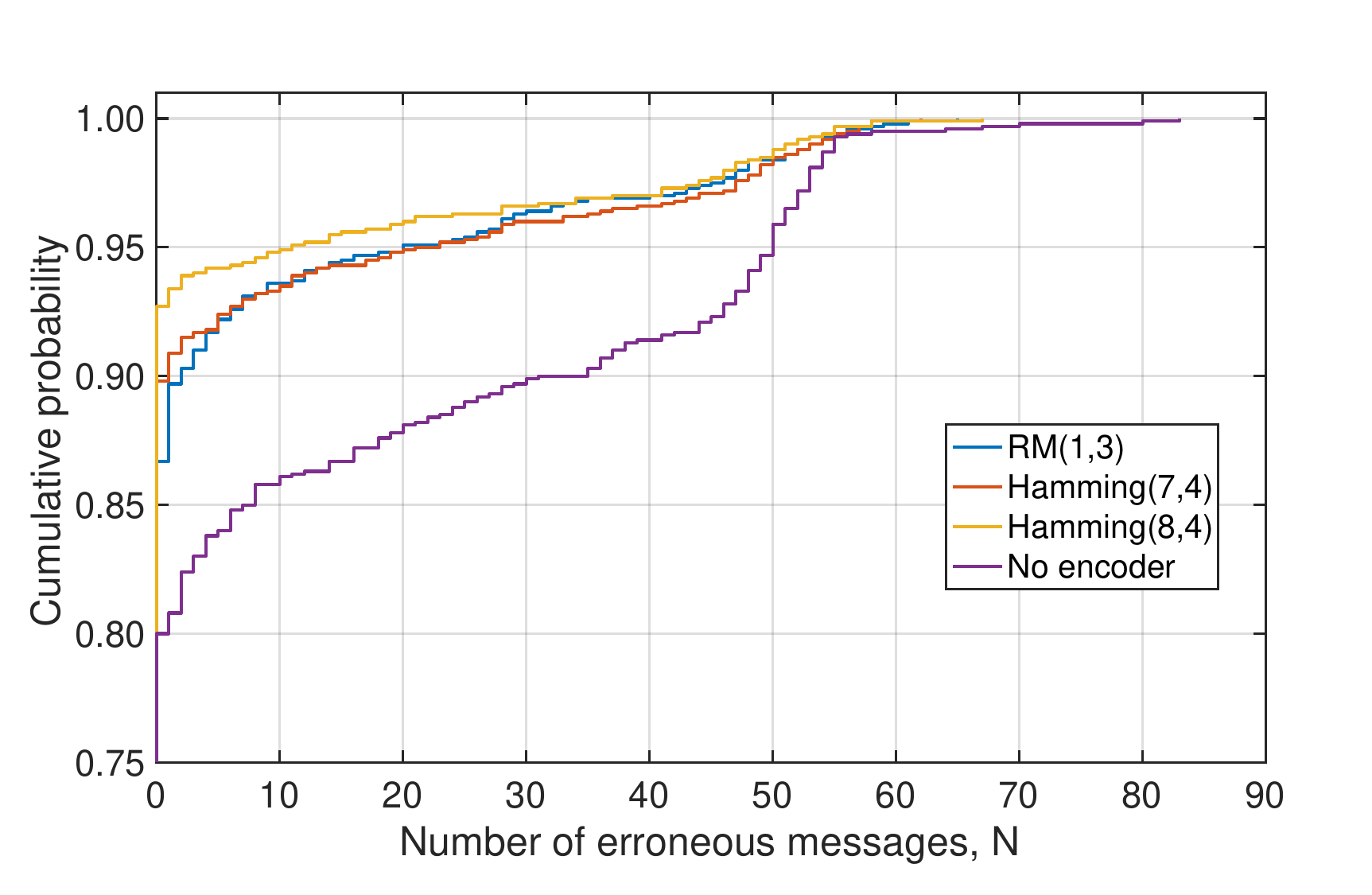}
	\caption{CDF representing the probability of receiving at most $N$ erroneous messages out of 100 transmissions. Each message was transmitted 1,000 times under independently sampled process variations, with each iteration incorporating up to ±20\% variation in process parameters (set by JoSIM simulator).}
	\label{fig:CDF_plot_comparison_RM_Hamming}
\end{figure}

\cref{fig:CDF_plot_comparison_RM_Hamming} presents the cumulative distribution function (CDF) of receiving at most $N$ erroneous messages within a sequence of 100 consecutive transmissions, evaluated for each error-correction coding scheme. 
Additionally, a `no encoder' data is added that corresponds to a 4-bit communication without any encoders and decoders. 
In \cref{fig:CDF_plot_comparison_RM_Hamming}, 100 random messages are sent through an encoder circuit with $\pm$20\% PPV spread. 
This setup is repeated 1000 times to achieve sufficient coverage of PPV values. 
Note that each iteration can be viewed as a distinct fabricated chip with specific circuit parameter values. 
Based on the observed data in \cref{fig:CDF_plot_comparison_RM_Hamming}, the probability of having zero errors in 100 decoded messages is 80.0\% without an encoder, and increases to 86.7\% for RM(1,3), 89.8\% for Hamming(7,4), and 92.7\% for Hamming(8,4) code encoders. 
Therefore, the Hamming(8,4) code provides a better trade-off in terms of error-correction capability as compared to other encoders.

Based on the the worst- and best-case comparisons of the error-correction codes presented in \cref{tab:lwc}, one can argue that RM(1,3) code is expected to perform slightly better than Hamming(8,4). 
However, from the circuit analysis presented in \cref{table:power_area_comparison}, it can be observed that RM(1,3) code encoder has a larger number of JJs as compared to the Hamming(8,4) code encoder. 
The larger number of JJs could result in a higher probability of circuit failure due to PPV, which has been confirmed in \cref{fig:CDF_plot_comparison_RM_Hamming}. 
However, having a simpler encoder circuit (\textit{e.g.}, Hamming(7,4) with the lowest number of JJs in \cref{table:power_area_comparison}) does not guarantee the most optimal performance. 
Therefore, there is a trade-off between the theoretical complexity of error-correction code and the physical size of the circuit-level implementation.







\section{Conclusion}\label{conclusion}
In this paper, three lightweight error-correction code encoders are studied for superconducting digital electronics applications. 
In particular, Hamming(7,4), Hamming(8,4), and RM(1,3) codes are implemented at the circuit level using SFQ logic gates. 
To evaluate the performance of these encoders, a simulation framework comprising JoSIM SPICE simulator and MATLAB tools has been proposed. 
The effect of PPV, one of the primary sources non-idealities in superconducting circuit fabrication, on the performance of error-correction encoders is evaluated through extensive simulations. 
Among the encoders tested, the Hamming(8,4) code demonstrated the highest probability of transmitting a message without bit errors. 
The trade-off between the theoretical complexity and physical size of error-correction code encoders is discussed.


\bibliographystyle{./bibliography/IEEEtran}
\bibliography{./bibliography/IEEEabrv,./bibliography/IEEEexample}

\end{document}